\pdfoutput=1
\documentclass[a4paper]{cas-dc}

\usepackage{caption}
\usepackage{amsmath,amssymb,amsfonts}
\usepackage{amsthm}
\usepackage{mathrsfs}
\usepackage[title]{appendix}
\usepackage{xcolor}
\usepackage{textcomp}
\usepackage{manyfoot}
\usepackage{algorithm}
\usepackage{algorithmicx}
\usepackage{algpseudocode}
\usepackage{listings}
\usepackage{dcolumn}
\usepackage{bm}
\usepackage{CJK}
\usepackage{array}
\usepackage{subfigure}
\usepackage{soul}
\usepackage{color}
\usepackage[justification=centering]{caption}
\usepackage{amsmath}
\usepackage[group-separator={,},group-minimum-digits=3]{siunitx}
\usepackage[sort,compress,numbers]{natbib}
\usepackage{booktabs}

\usepackage{hyperref}
\usepackage{multirow}
\usepackage{graphicx}
\usepackage{tabularx}
\usepackage{float}

\begin{document}
\shortauthors{P.~Zhang et al.}
\shorttitle{
Measurement of cosmic-ray muon flux at CJPL-II
    }
\title[mode = title]{
Measurement of cosmic-ray muon flux at CJPL-II
	}

\author[1]{P.~Zhang}
\author[1]{H.~Ma}
\author[1]{W.~Dai}
\author[1]{M.~Jing}
\author[1]{L.~Yang}
\author[1]{Q.~Yue}

\author[1]{Z.~Zeng}[orcid=0000-0003-1243-7675]
\cormark[1]
\cortext[1]{Corresponding author: zengzhi@tsinghua.edu.cn}

\author[1, 2]{J.~Cheng}
\cormark[2]
\cortext[2]{Corresponding author: chengjp@tsinghua.edu.cn}

\address[1]{Key Laboratory of Particle and Radiation Imaging (Ministry of Education) and Department of Engineering Physics, Tsinghua University, Beijing 100084, China}
\address[2]{School of Physics and Astronomy, Beijing Normal University, Beijing 100875, China}

\begin{abstract}
In China Jinping Underground Laboratory (CJPL),
the deepest and largest underground laboratory globally,
the cosmic-ray muon flux is significantly reduced due to the substantial shielding provided
by the overlying mountain. From 2016 to 2020, we measured the
muon flux in the second phase of CJPL (CJPL-II)
with a plastic scintillator muon telescope system, detecting 161 muon events
over an effective live time of 1098 days.
The detection efficiency was obtained by
simulating the underground muon energy and angular distributions
and the telescope system's response to underground muons.
The cosmic-ray muon flux is determined to be
(3.03 $\pm$ 0.24 (stat) $\pm$ 0.18 (sys)) $\times$ 10$^{-10}$ cm$^{-2}$s$^{-1}$,
which is the lowest among underground laboratories worldwide.
\end{abstract}


\maketitle

\section{Introduction}\label{sec1}
China Jinping Underground Laboratory (CJPL), located within the 17.5
km long traffic tunnel of the Jinping Mountain
in Sichuan Province, China, is shielded by
approximately \SI{2400}{\m} of vertical rock coverage
\cite{wu_measurement_2013, cheng_china_2017}.
This extensive rock shielding significantly reduces the cosmic-ray muon flux,
thereby minimizing the muon and muon-induced radiation backgrounds
\cite{zeng_characterization_2017,zeng_evaluation_2020,guo_muon_2021}
and making CJPL an ideal site for rare-event experiments,
such as direct dark matter detection and
neutrinoless double-$\beta$ decay experiments \cite{cheng_china_2017}.

The first phase of the laboratory (CJPL-I) began construction in 2009
and officially became operational in 2010,
with a total space of approximately \SI{4000}{\cubic\m} \cite{cheng_china_2017}.
In 2013, using a plastic scintillator muon telescope system,
the cosmic-ray muon flux at CJPL-I was measured to be
(2.0 $\pm$ 0.4) $\times$ 10$^{-10}$ cm$^{-2}$s$^{-1}$,
without efficiency correction to the detector angular acceptance
\cite{wu_measurement_2013}.
In 2021, the efficiency-corrected cosmic-ray muon flux at CJPL-I was determined to be
(3.53 $\pm$ 0.23) $\times$ 10$^{-10}$ cm$^{-2}$s$^{-1}$ based on data from a liquid
scintillator prototype system for solar neutrino detection \cite{guo_muon_2021}.

The second phase (CJPL-II), located about \SI{500}{\m} west of CJPL-I
along the same traffic tunnel, finished the civil engineering
and decoration in 2024 with a capacity of \SI{300000}{\cubic\m}
\cite{cheng_china_2017}.
This study employed the same plastic scintillator muon telescope system\cite{wu_measurement_2013}
to measure the cosmic-ray muon flux in the hall C2 of CJPL-II.

This article is structured as follows:
Section \ref{sec2} details the experiment,
the simulation of the underground muon distributions
and the telescope system's response to underground muons.
Section \ref{sec3} presents the results of the experiment and simulation.
Conclusions are drawn in Section \ref{sec4}.

\section{Experiment and methods}\label{sec2}
\begin{figure*}[!htb]
    \centering
    \includegraphics[width=1\hsize]
    {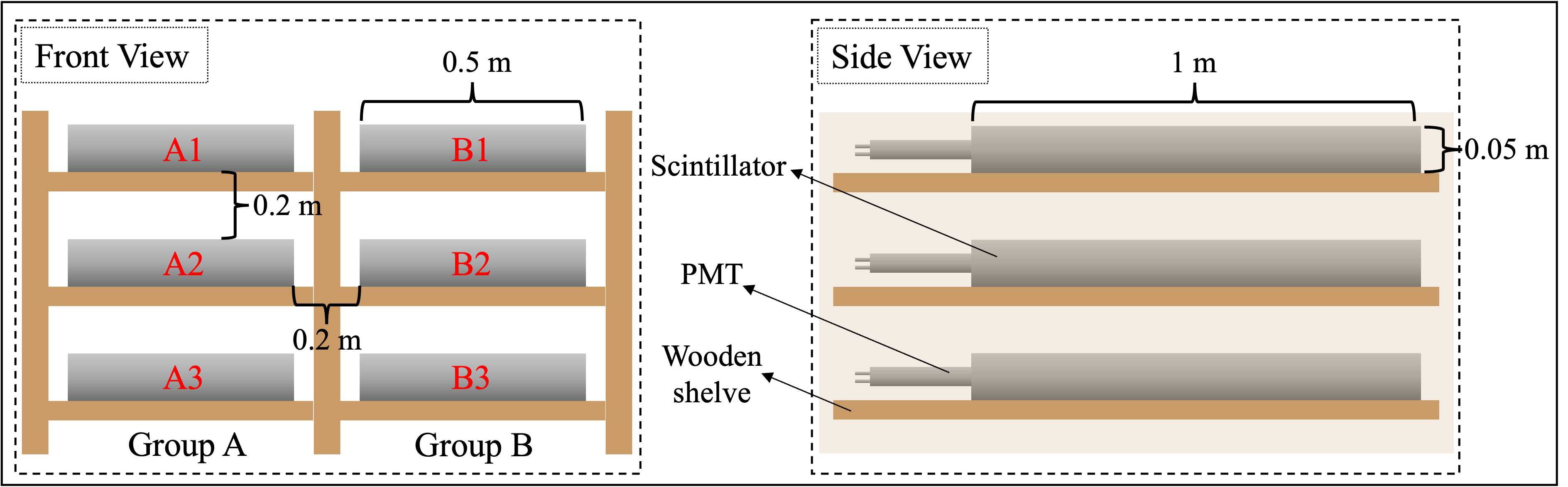}
    \caption{Schematic of the muon telescope system}
    \label{fig1}
\end{figure*}

\begin{figure*}[htb!]
    \centering
    \includegraphics[width=1\hsize]
    {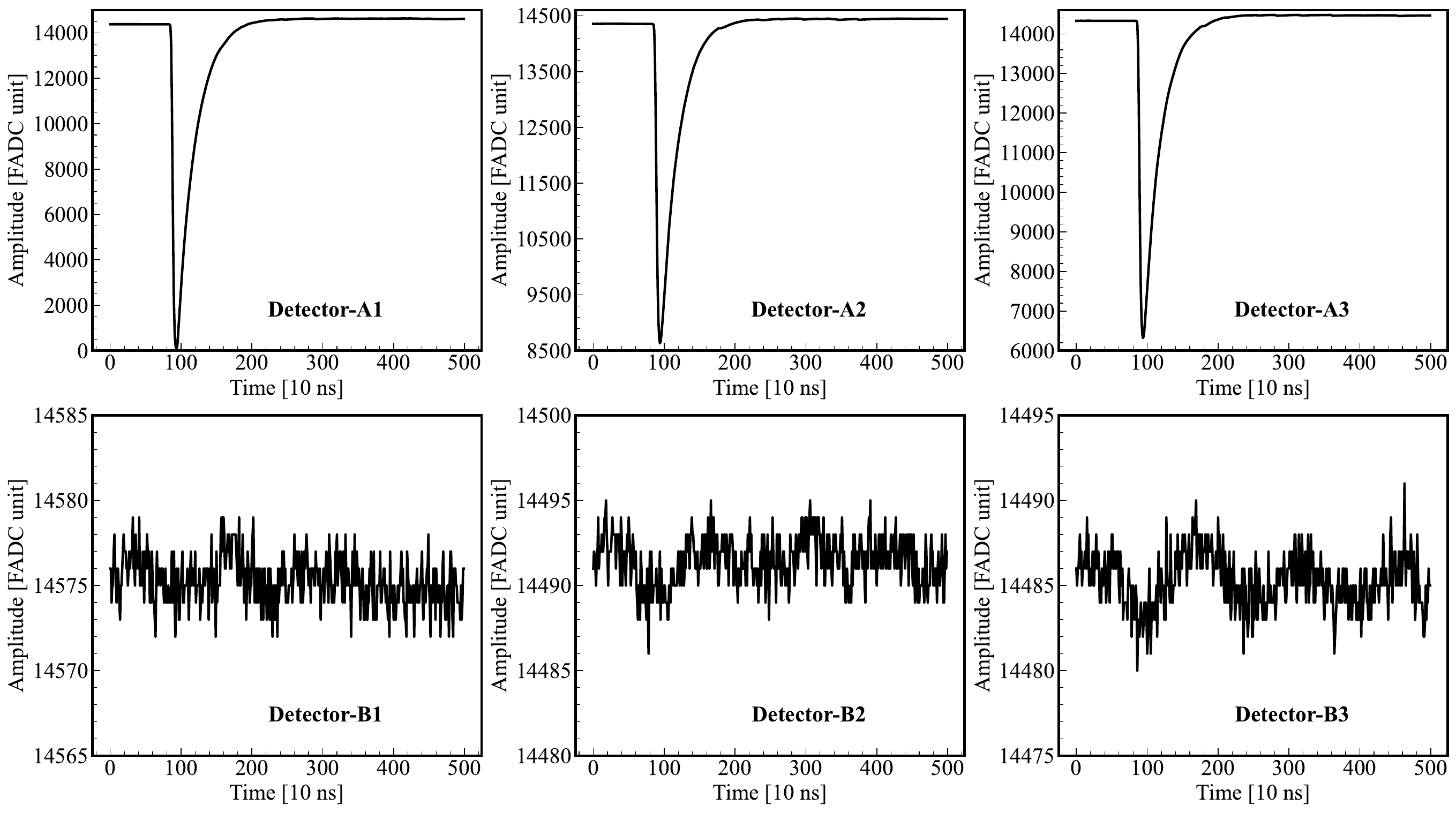}
    \caption{Detector waveforms when a muon passed
    all three detectors of group A.
    The deposited energy in the A1, A2 and A3 detectors are
    19.3, 9.7 and 12.3 MeV.
    }
    \label{fig2}
\end{figure*}

From October 2016 to December 2020, we conducted
a measurement of the cosmic-ray muon flux in the hall C2 of CJPL-II using a
plastic scintillator muon telescope system.
The telescope system, depicted in Figure \ref{fig1},
comprises two groups of plastic scintillators, each containing three detectors.
The size of each detector is 1 m $\times$ 0.5 m $\times$ 0.05 m.
The distances between neighboring detectors (vertically/horizontally) are both about 0.2 m \cite{wu_measurement_2013}.
With ultrahigh energy and strong penetration capability,
a cosmic-ray muon could pass through all three detectors in a group,
resulting in triple coincidence signals with large amplitudes.
The muon candidates were selected after the data quality check
(remove data batches with abnormal trigger rate)
, the rise/fall time restrain and the amplitude restrain \cite{wu_measurement_2013}.
For simplification, we didn't apply anti-coincidence between two groups
in the event selection.
Figure \ref{fig2} shows the detector waveforms of a muon candidate.
The measured muon flux $\phi_{\text{m}}$ (efficiency-uncorrected) is calculated as follows:
\begin{equation}
\phi_{\text{m}} = \frac{N_\mu}{S_{\text{d}} \cdot T}
\label{eq.phim}
\end{equation}
where $N_\mu$ is the number of the selected muon candidates,
$S_{\text{d}}$ (0.5 m$^2$) is
the upper surface area of the detector,
and $T$ is the experimental live time.

On the one hand,
only part of the muons passing through the top detectors could
be selected as the muon candidates.
On the other hand,
the selected muon candidates may be contaminated
by high-energy muon shower events, which are generated
when muons pass through rock around the laboratory.
Therefore, the actual integrated muon flux $\phi$ (efficiency-corrected) is
related to the measured muon flux $\phi_{\text{m}}$
with the overall detection efficiency $\epsilon$:
\begin{equation}
\phi = \frac{\phi_{\text{m}}}{\epsilon}
\end{equation}

To calculate the detection efficiency,
we firstly need to determine the energy and angular distributions of 
underground muons by simulating the transportation of high-energy muons
through the mountain above.
Then, the detection efficiency is calculated by simulating
the telescope system's response to underground muons.

\begin{figure}[htb!]
    \centering
    \includegraphics[width=1\hsize]
    {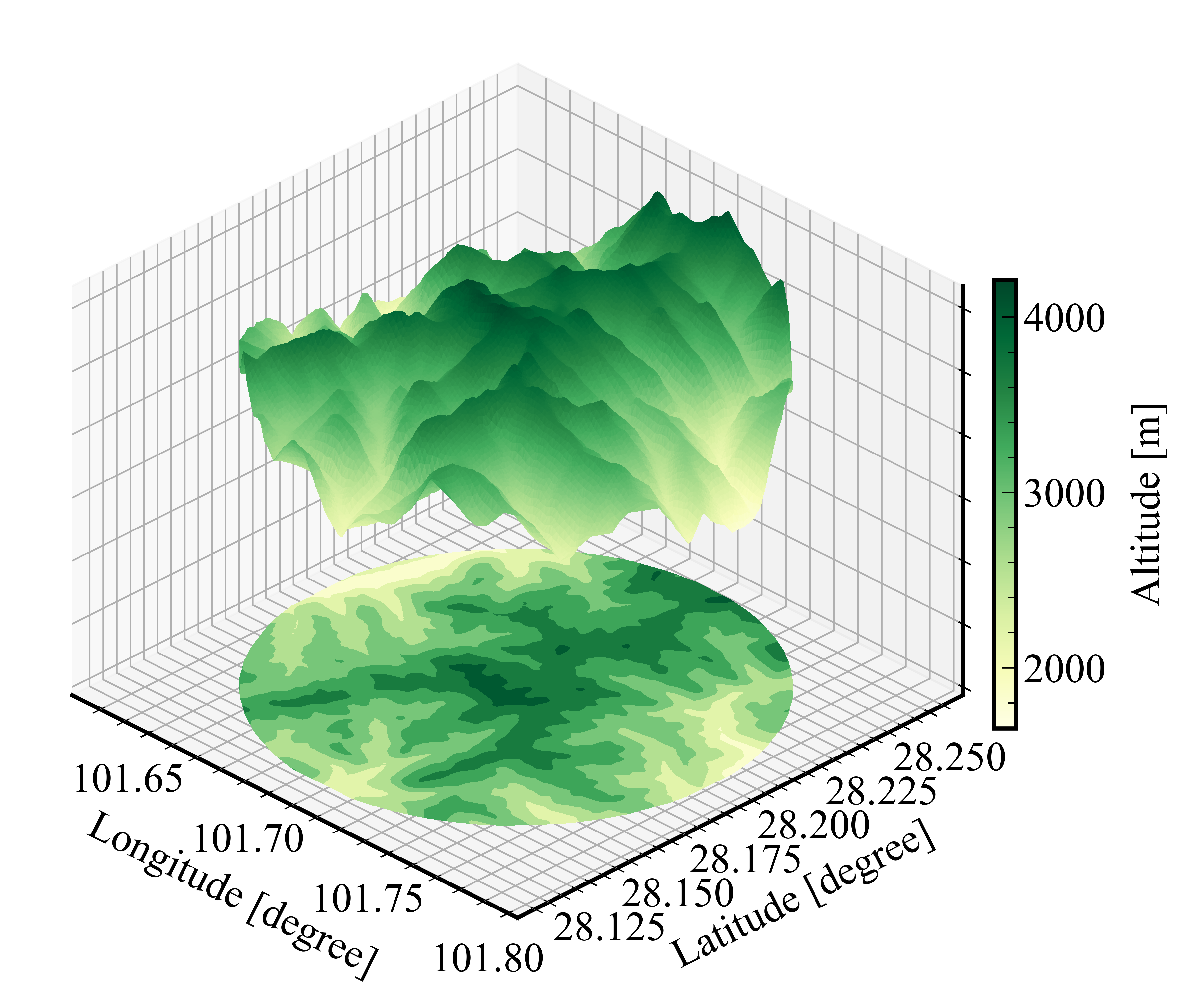}
    \caption{\label{fig3}
    Geometry model of the Jinping Mountain
    }
\end{figure}

CJPL is located beneath the Jinping Mountain,
at an altitude of approximately \SI{1600}{m},
with a rock overburden of about \SI{2400}{m}.
The Google Earth program\cite{noauthor_google_2024}
was utilized to obtain the latitude, longitude,
and elevation of \SI{18529} points on the mountain surface
within an 8 km radius circle centered at the CJPL-II,
with a grid size of 0.001 degree $\times$ 0.001 degree
(latitude $\times$ longitude). 
The geometry model of the Jinping Mountain is
visualized in Figure \ref{fig3}.
We employed a modified Gaisser Formula
to sample the incident high-energy muons \cite{guan_parametrization_2015}.
The angular dependence of muon energy distribution is considered.
For each muon, the depth of rock
along the muon's path to the laboratory was calculated.
Then, we simulated the transportation of each muon
using the MUSIC program \cite{kudryavtsev_muon_2009},
which was widely used in the muon flux simulations for underground laboratories.
The average density and the composition of rock
follow the setting in Ref. \cite{guo_muon_2021}.
The muon energy and angular distributions underground were obtained
by analyzing the transportation results.

\begin{figure}[htb!]
    \centering
    \includegraphics[width=1\hsize]
    {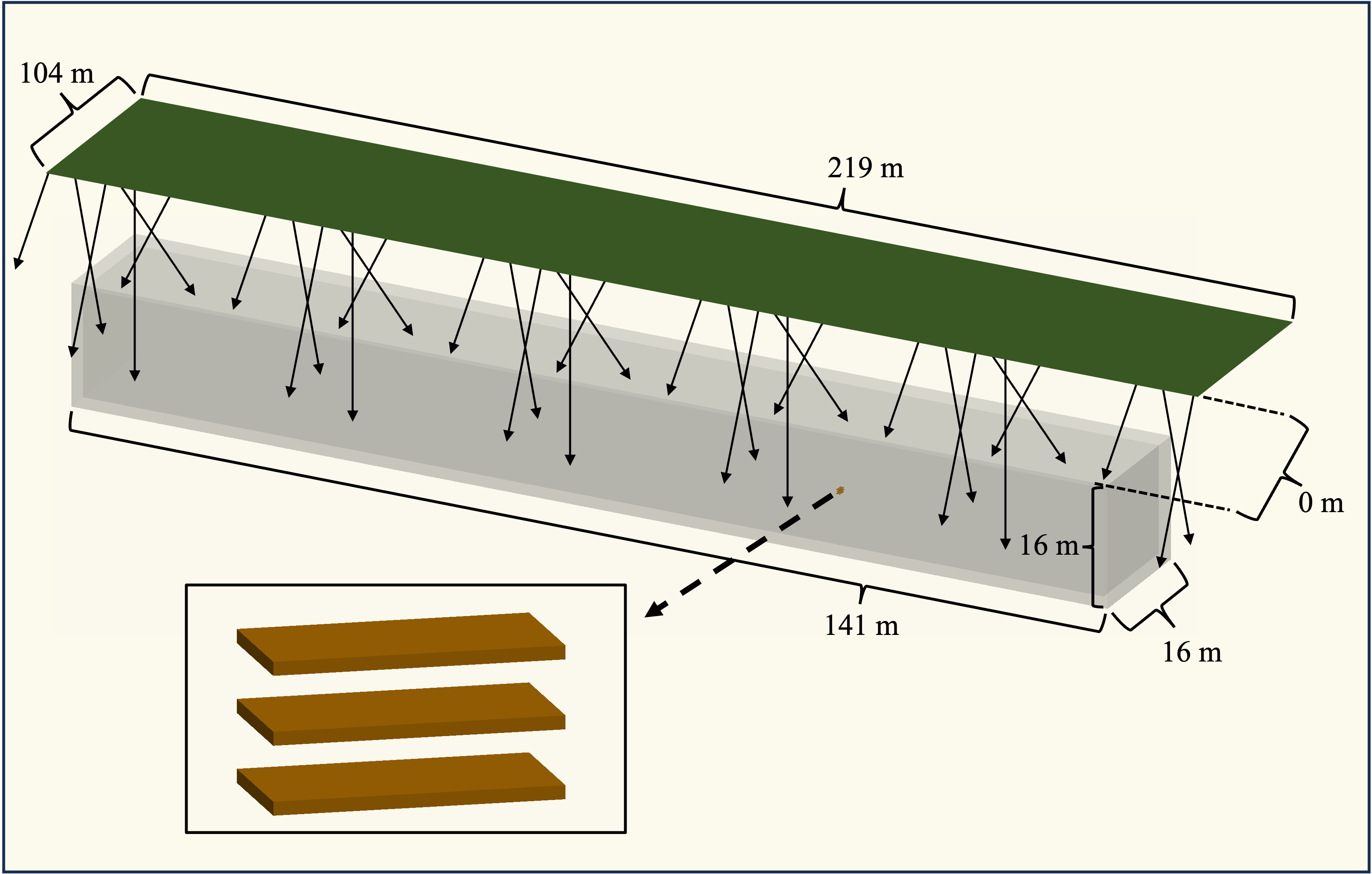}
    \caption{
    Visualization of the telescope system (zoomed in) at the hall C2 of CJPL-II.
    The green plane represents the sampling plane,
    which is in the same plane of the upper surface of the rock layer}.
    The arrows illustrated
    the tracks of the sampled underground muons.
    \label{fig4}
\end{figure}

We utilized the Geant4 program
\cite{allison_geant4_2006,allison_recent_2016,agostinelli_geant4simulation_2003}
to simulate the telescope's response to underground muons.
As illustrated in Figure \ref{fig4},
we constructed a geometry model in Geant4
with the internal size of the experimental hall set to
129 m $\times$ 14 m $\times$ 14 m.
We added 1 m thick rock around the hall to simulate
the electromagnetic shower initiated by muon,
following the setting in Ref. \cite{guo_muon_2021}.
Upon evaluation, 1 m thick rock is thick enough and
the result variation is within the margin of error when the depth is set bigger.
The density and composition of rock were set the same with 
those used in the muon transportation simulation.
With the simulated underground muon distributions,
muon events were sampled on a sampling plane
above the experimental hall.
The sampling plane was set big enough
so that the muons that could pass through the experimental hall
could be sampled from the sampling plane.

Let $N$ denote the total number of the muons simulated,
$N_{\text{a}}$ the number of the muons passing through all three detectors,
and $N_{\text{t}}$ the total number of events passing the energy threshold.
Thus, the angular acceptance $\epsilon_{\text{a}}$ and
the overal efficiency $\epsilon$ are calculated as follows:
\begin{gather}
\epsilon_{\text{a}} = \frac{N_{\text{a}} \cdot C_r}{N},
\label{eq.effang}
\epsilon = \frac{N_{\text{t}} \cdot C_r}{N}\\
C_r = \frac{S_{\text{s}}}{S_{\text{d}}}
\label{eq.cs}
\end{gather}

where $S_{\text{s}}$ (22750 m$^2$)
is the area of the sampling plane.
With the factor $C_r$ introduced,
the top detector upper surface plane
becomes the reference plane of the efficiencies.
The angular acceptance $\epsilon_{\text{a}}$ is equivalent
to the so called "solid angle correction factor",
which is defined as the ratio of  the number of the muons passing through all three detectors
over the number of the muons pasing through the top detector \cite{wu_measurement_2013}.

\section{Cosmic-ray muon flux at CJPL-II}\label{sec3}
\begin{figure}[htb!]
    \centering
    \includegraphics[width=1\hsize]
    {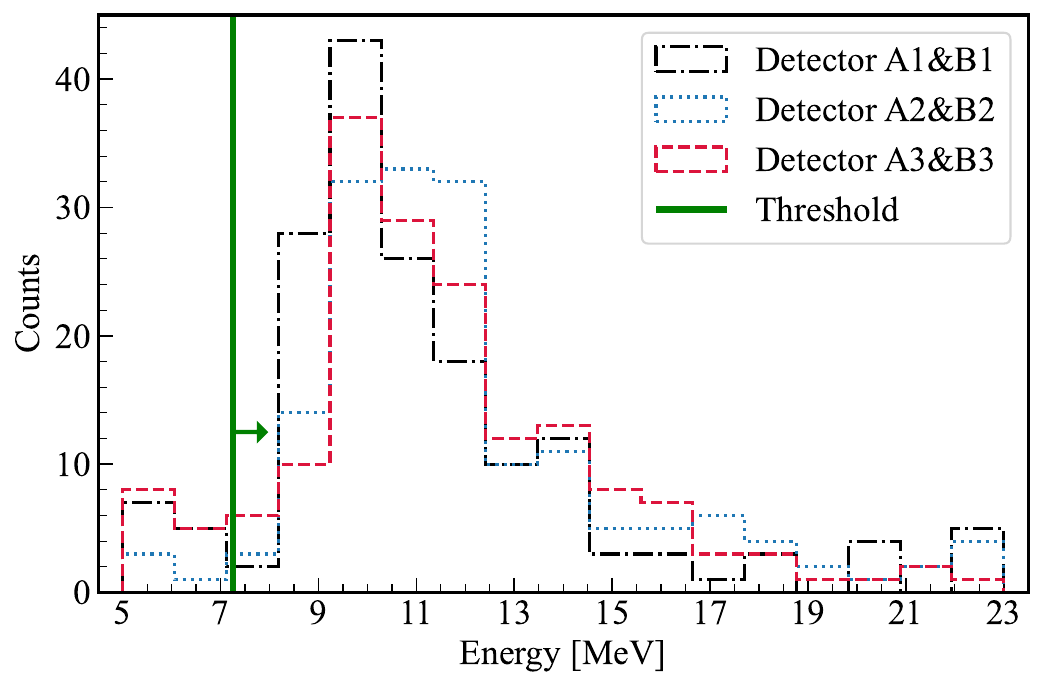}
    \caption{Energy spectra of triple coincidence signals}
    \label{fig5}
\end{figure}

After the data quality check,
8.5 days of data were discarded,
and then we obtained a dataset with an effective live time of 1098 days. 
The energy calibration was performed by
comparing the simulated energy distribution
and the measured amplitude distribution.
The systematic uncertainty of the obtained energy is about 3.1\%.
The energy threshold for event selection was set to 7.3 MeV,
which is the lower energy limit of the muon spectrum peak.
Figure \ref{fig5} presents the energy spectra 
of triple coincidence signals.
The measurement results are summarized in Table \ref{tab1}.
Among total 161 events,
there are 85 events from group A and 76 events from group B.
Besides, there are four pairs of coincidental events between group A and group B,
which corresponds to the expectation from simulation.

\begin{figure}[htb!]
    \centering
    \includegraphics[width=1\hsize]
    {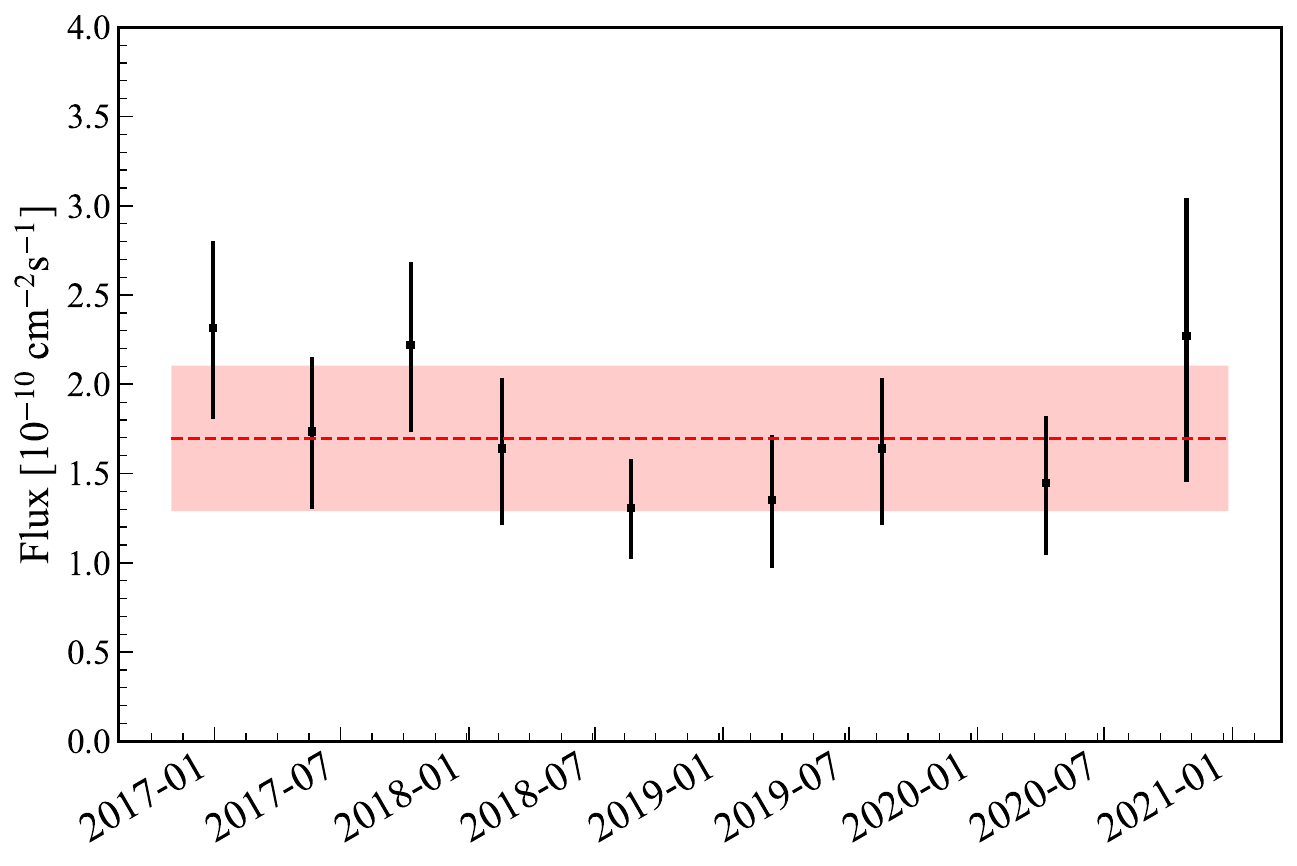}
    \caption{
    Measured cosmic-ray muon flux over time without efficiency correction
    }
    \label{fig6}
\end{figure}

Combining the data from two groups,
the measured muon flux $\phi_{\text{m}}$ was determined to be
(1.70 $\pm$ 0.13) $\times$ 10$^{-10}$ cm$^{-2}$s$^{-1}$,
while the same telescope system yielded a result of
(2.0 $\pm$ 0.4) $\times$ 10$^{-10}$ cm$^{-2}$s$^{-1}$
at CJPL-I in 2013 \cite{wu_measurement_2013}.
As shown in Figure \ref{fig6},
we grouped the data into four-month intervals and calculated
the flux for each interval.
In the shallower underground laboratories,
seasonal modulations of the muon flux were observed
\cite{agostini_modulations_2019,prihtiadi_measurement_2021}.
In this study, no modulation effect of
the muon flux over time was observed
(In the flat distribution assumption, $\chi^2$/d.o.f = 19.3/16, p-value = 0.25),
due to significant statistical fluctuations.

\begin{table*}[t]
    \centering
    \caption{
    Results of cosmic-ray muon flux measurement
    }
    \label{tab1}
    \begin{tabular}{cccc}
      \toprule
      Detector group & Live time (days) & Number of muon events & Muon flux (cm$^{-2}$s$^{-1}$) \\
      \hline
      A & 1098 & 85 & $(1.80 \pm 0.19) \times 10^{-10}$ \\
      B & 1098 & 76 & $(1.60 \pm 0.18) \times 10^{-10}$ \\
      Total & 1098 & 161 & $(1.70 \pm 0.13) \times 10^{-10}$ \\
      \bottomrule
    \end{tabular}
\end{table*}

\begin{figure}[htb!]
    \centering
    \includegraphics[width=1\hsize]
    {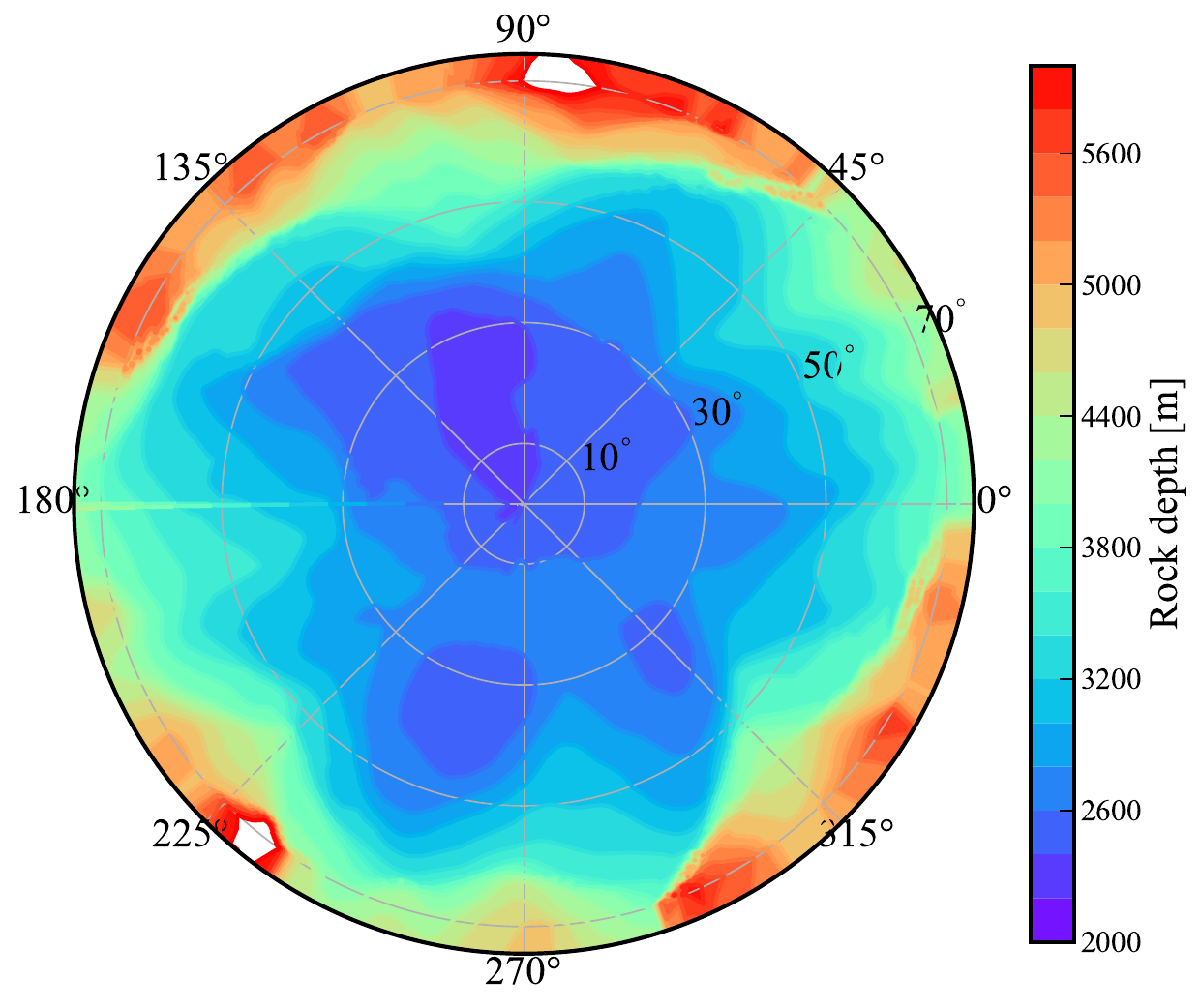}
    \caption{
    Distribution of depth of rock that the incident muons traverse.
    The radial coordinate represents the zenith angle,
    and the angular coordinate represents the azimuth angle
    (degree 0 corresponds to the east direction).
    }
    \label{fig7}
\end{figure}

\begin{figure}[htb!]
    \centering
    \includegraphics[width=1\hsize]
    {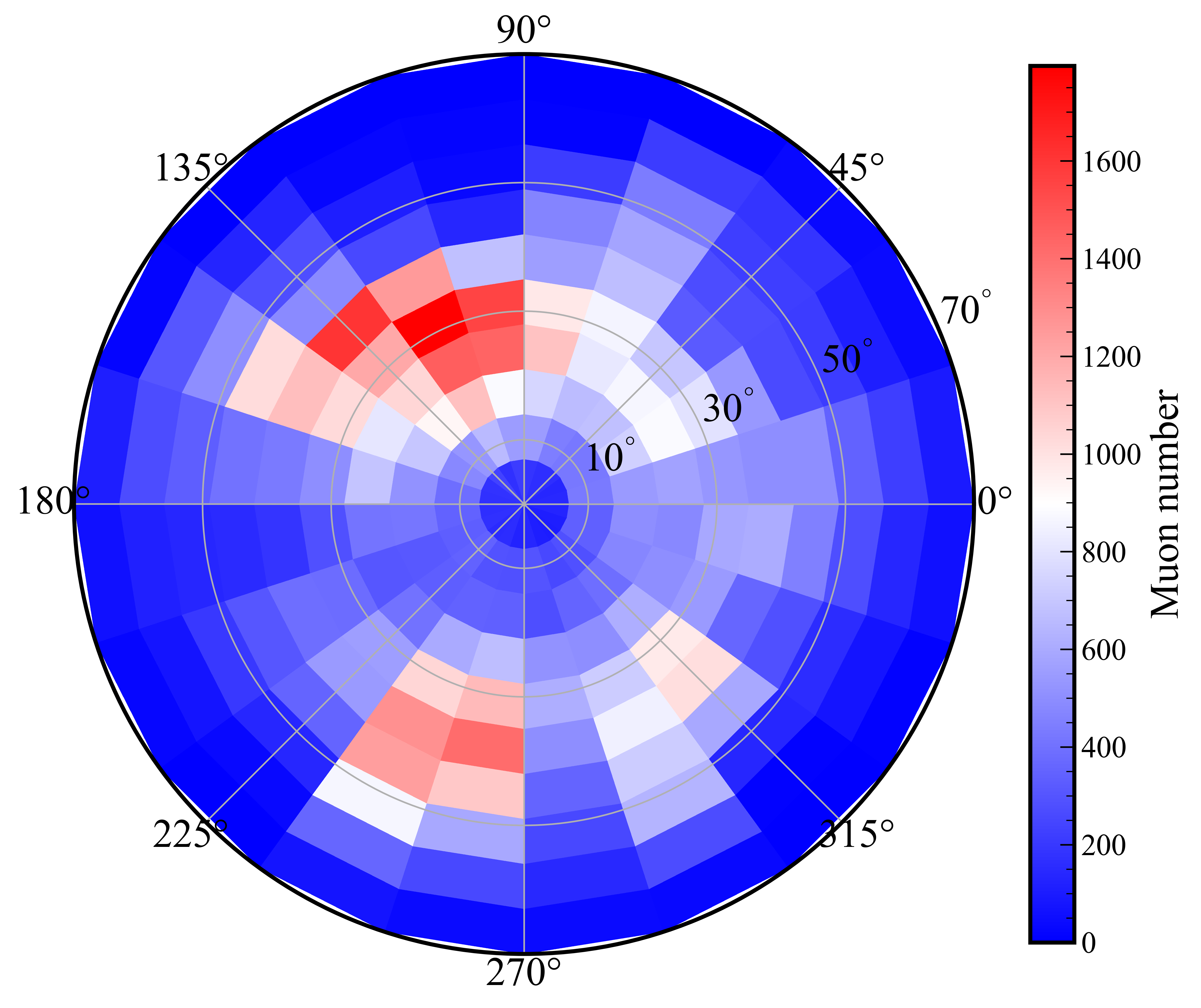}
    \caption{
    Simulated angular distribution of the underground muons in polar coordinates
    The radial coordinate represents the zenith angle,
    and the angular coordinate represents the azimuth angle
    (degree 0 corresponds to the east direction).
    }
    \label{fig8}
\end{figure}

\begin{figure}[htb!]
    \centering
    \includegraphics[width=1\hsize]
    {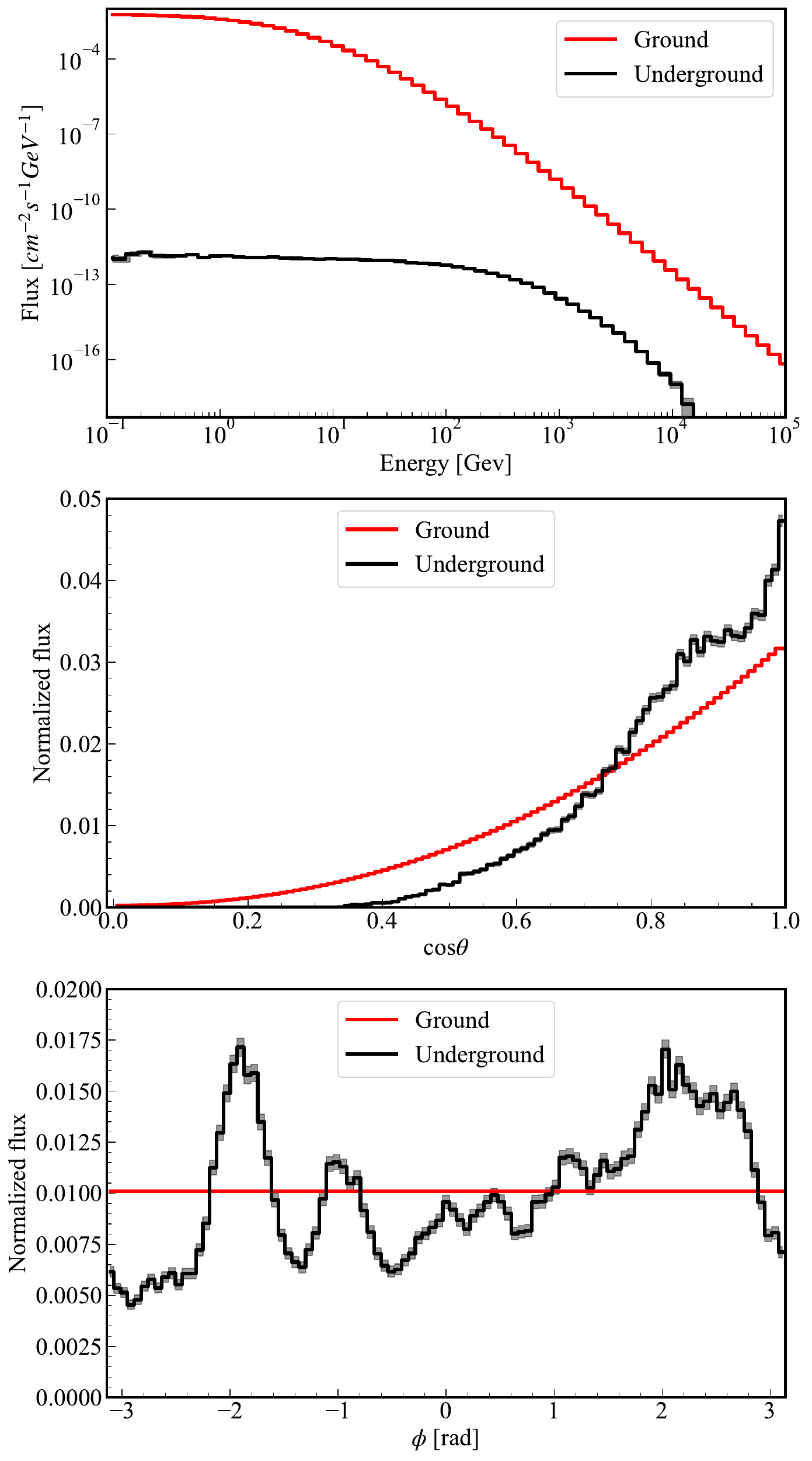}
    \caption{
    \label{fig9}
    Energy and angular distributions of muons on the ground and underground.
    }
\end{figure}

To calculate the energy and angular distributions of underground muons,
a total of $2 \times 10^7$ high-energy incident muons were simulated.
The distribution of depth of rock that the incident muons traverse
is shown in polar coordinates in Figure \ref{fig7}.
The angular distribution of the underground muons
is shown in Figure \ref{fig8}.
Simulation results indicates that
the angular distribution of the surviving muons
is consistent with the distribution of the depth
of rock that the incident muons traverse.
A significant portion of the muon flux at CJPL-II
originates from the northwest and southwest,
with zenith angles of 20-40 degrees,
where the rock layer is thinner.
The energy and angular distributions of muons on the ground
and underground are compared in Figure \ref{fig9}.
The mountain reduces the flux of 4 GeV muons by approximately 9
orders of magnitude and the integral flux by about 8 orders of
magnitude. The average energy of the underground muons is 359.6 GeV,
which is close to the results in Ref. \cite{zeng_evaluation_2020,guo_muon_2021}.

With the simulated underground muon distributions,
we sampled $1 \times 10^9$ underground muons to
simulate the response of the telescope system.
The muons with zenith angle above 70 degrees
are ignored in the simulation because they contribute
less than 0.04\% of the detector counts.
The angular acceptance $\epsilon_\text{a}$
and the overall efficiency $\epsilon$
were determined to be
0.510 $\pm$ 0.005 (stat) $\pm$ 0.046 (sys)
and 0.562 $\pm$ 0.005 (stat) $\pm$ 0.033 (sys).
The angular acceptance is bigger than the value
of 0.339 on the ground \cite{wu_measurement_2013},
because the angular distribution of the muons at CJPL-II
is more vertical than that on the ground, as shown in Figure. \ref{fig9}.

The uncertainty sources of the overall efficiency are summarized in
Table \ref{tab2}. The systematic uncertainty of the efficiency arises
from three sources: the mountain geometry model, the density of rock and the thickness
of rock around the experimental hall.
We evaluated the uncertainty from the mountain geometry model by shifting it by $\pm$ 100 m in the north-south,
east-west, and up-down directions, taking the maximum deviation as a
part of systematic uncertainty. The uncertainty from the rock density
was evaluated by considering variations of $\pm$ 0.1 g/cm$^3$,
while the uncertainty from the rock thickness was assessed
by considering variations of $\pm$ 0.5 m.

\begin{table}[t]
    \centering
    \caption{
    Summary of uncertainties of the overall efficiency
    }
    \label{tab2}
    \begin{tabular}{cc}
      \toprule
      Uncertainty source & Uncertainty \\
      \hline
      Systematic (Mountain geometry) & 0.019 \\
      Systematic (Rock density) & 0.022 \\
      Systematic (Rock thickness) & 0.016 \\
      Statistical & 0.005 \\
      Total & 0.033 \\
      \bottomrule
    \end{tabular}
\end{table}

With the measured muon flux $\phi_\text{m}$
and the overall efficiency $\epsilon$ calculated above,
we derived that the muon flux $\phi$ at CJPL-II is
(3.03 $\pm$ 0.24 (stat) $\pm$ 0.18 (sys)) $\times$ 10$^{-10}$ cm$^{-2}$s$^{-1}$.
It is slightly lower than those of SNOLab \cite{aharmim_measurement_2009}
and CJPL-I \cite{guo_muon_2021}.

\section{Conclusion}\label{sec4}
We present a study of the cosmic-ray muon flux measurement
using a plastic scintillator muon telescope system at CJPL-II.
To calculate the detection efficiency,
we first simulated the transportation of muons through the Jinping mountain
and obtained the energy and angular distributions of the underground muons.
Results indicate that a significant contribution to the muon flux
at CJPL-II originates from the northwest and southwest,
with zenith angle of 20-40 degrees.
Then, using the simulated underground muon distributions,
we simulated the telescope system's response
to underground muons and
derived the detection efficiency.
Finally, the cosmic-ray muon flux at CJPL-II
was determined to be (3.03 $\pm$ 0.24 (stat) $\pm$ 0.18 (sys)) $\times$ 10$^{-10}$ cm$^{-2}$s$^{-1}$,
which is the lowest among underground laboratories worldwide.

\section*{Acknowledgments}
This work was supported by
the National Natural Science Foundation of China
(Grant No. 12425507 and No. 12175112)
and
the National Key Research and Development Program of China
(Grant No. 2023YFA1607101 and No. 2022YFA1604701).
The authors would like to thank Dr. Weihe Zeng
for helpful discussion on the MUSIC simulation.

\bibliography{ref.bib}
\bibliographystyle{JHEP}

\end{document}